\documentstyle[12pt,psfig]{article}
\def\reference{\parskip 0pt\par\noindent\hangindent 0.5 truecm}
\def\kms{km ${\rm s}^{-1}$}

\def\simlt{\lower.5ex\hbox{$\; \buildrel < \over \sim \;$}}
\def\simgt{\lower.5ex\hbox{$\; \buildrel > \over \sim \;$}}
\begin{document}
%
% Title
% Capitalise the title normally - do not use ALL CAPS.
%
\title{The Optical Counterpart of  
the X-ray Transient \\RX J0117.6$-$7330: Spectroscopy and Photometry}
%

% Authors
% Here comes the author(s) of the paper. Please add the appropriate author
% names for your paper and indicate within the $^...$ the number(s)
% which corresponds to the institute(s) of each author. In this example
% the second author has two institutional affiliations.
% Add or remove authors as required, maintaining the \and syntax between
% each author, but no \and after the last author.
% **** IMPORTANT: Leave the closing curly bracket line as is. ******

\author{Roberto Soria $^{1}$ 
} % IMPORTANT: leave this curly bracket as the first character of this line.

% Date - leave this blank.
\date{}
\maketitle

% Institutions
% Here fill in your institute name(s) and address(es)
% The number in $^...$ indicates the author number.  For example
{\center
$^1$ Mount Stromlo Observatory, Private Bag, Weston Creek P.O., 2611, ACT, 
Australia\\roberto@mso.anu.edu.au\\[3mm]
}

% Abstract
% Simply place your abstract between the \begin{abstract} and
% \end{abstract} commands.
%
\begin{abstract}
% Place the abstract here.

We conducted spectroscopic and photometric observations of the 
optical companion of the X-ray transient RX J0117.6$-$7330 in 
the Small Magellanic Cloud, during a quiescent state. 
The primary star is identified as 
a B$0.5$\,IIIe with a mass $M_* = (18 \pm 2) M_{\odot}$ and bolometric 
magnitude $M_{\mbox{{\scriptsize bol}}} = -7.4 \pm 0.2$. 
The main spectral features are strong H$\alpha$ emission, H$\beta$ and 
H$\gamma$ emission cores with absorption wings, 
and narrow He{\small{ I}} and O{\small{ II}} absorption lines. 
Equivalent width and full width at half maximum of the main lines are 
listed. The average systemic velocity over our observing run is 
$v_r = (184 \pm 4)$ \kms; measurements over a longer period of 
time are needed to 
determine the binary period and the $K$ velocity of the 
primary. We determine a projected rotational velocity 
$v \sin i = (145 \pm 10)$ \kms for the Be star, and we deduce that 
the inclination angle of the system is $i = (21 \pm 3)\deg$.

\end{abstract}

{\bf Keywords:}
% Place keywords here.  PASA uses the standard list of subject 
% headings adopted by The Astrophysical Journal and available from URL:
%   http://www.noao.edu/apj/keywords96.html

 galaxies: Magellanic Clouds---stars: individual 
(RX J0117.6$-$7330)---X-ray: binaries

% A formatting command to add space between the author list and the body
% of the paper when printed. This spacing may be changed as desired.
\bigskip

%
% Body of paper
%

\section{Introduction}

% Place contents of first section here.

The X-ray transient RX J$0117.6-7330$ was discovered by the Position 
Sensitive Proportional Counter on board ROSAT on October 1-2, 
1992 (Clarke, Remillard \& Woo 1996, 1997); it is located approximately 
$5'$ south-east of the X-ray binary SMC X-$1$. Its optical counterpart 
was observed by 
Charles, Southwell \& O'Donoghue (1996), and by Clarke et al.\  
(1997), and has the characteristics of a Be star; the system is therefore 
likely to be a High-Mass X-ray Binary (HMXB).

Systems with a Be-type primary are the largest group among the HMXBs, 
both in the Galaxy and in the Magellanic Clouds (van Paradijs 1995; 
Kahabka \& Pietsch 1996). In these systems, X-ray outbursts lasting 
from weeks to months are caused by sudden enhancements of the 
equatorial mass loss of the Be star; during an active state, 
a modulation corresponding to the orbital period is often observed, 
with increased X-ray emission when the compact object transits 
near periastron in an eccentric orbit 
(van den Heuvel \& Rappaport 1987). Active states are separated 
by longer inactive periods, often lasting several years.

The nature of the compact object in RX J$0117.6-7330$ is not yet known: 
its soft X-ray spectrum, its long decay time after the 1992 outburst and 
the absence of pulsations make this system a possible black-hole candidate 
(Clarke et al.\ 1997). A common way to determine 
whether the compact object in an X-ray binary can be a neutron star 
is to deduce its 
mass function from the radial velocity amplitude of the primary; such 
measurements are more difficult in HMXBs owing to the low velocity 
expected for the primary and to the long binary period (tens to hundreds 
of days).

\section{Observations and Data Analysis}

% Place contents of next section here.

We observed the optical counterpart of RX J0117.6$-$7330 from August 20
to August 23, 1998, simultaneously with the 40inch telescope (photometry) 
and the ANU 2.3m telescope (spectroscopy) at Siding Spring Observatory.
Conditions were photometric during the first half of the first night 
and on the last night.

\subsection{High-resolution Optical Spectroscopy}

Optical spectra of the primary were obtained with the 
Double Beam Spectrograph on the 2.3m ANU 
telescope at Siding Spring Observatory, with 1200 grooves/mm gratings 
for both the blue ($4150-5115$ \AA) and the red ($6200-7140$ \AA) 
spectral regions (resolution 1.2 \AA\ FWHM); the detectors used were 
SITe $1752 \times 532$ CCDs in both arms of the spectrograph. 

Figure 1 and Figure 2 show the average of seven 600s spectra 
taken on August 20 in photometric conditions, for the blue and the 
red part of the spectrum respectively. Atmospheric absorption bands at 
$\lambda > 6860$ \AA \ have been removed from the red spectrum using 
the spectra of the calibration star LTT7379. Wavelengths are heliocentric.

\begin{figure}
\begin{center}
\centerline{\psfig{file=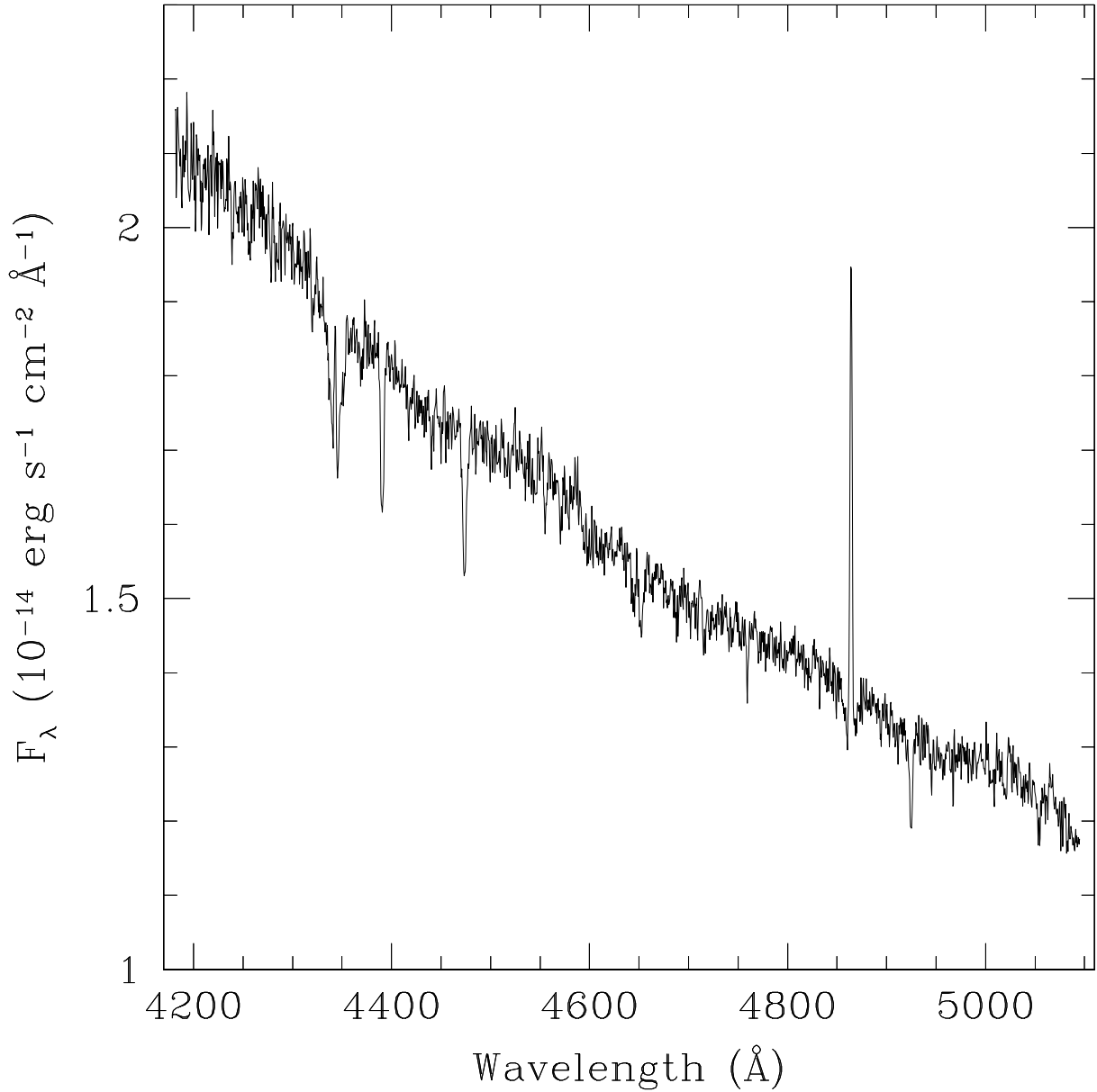,width=17cm}}
\caption{Flux-calibrated blue spectrum of the optical counterpart 
of RX J0117.6$-$7330, taken on August 20, 1998 from the ANU 2.3m telescope 
at Siding Spring Observatory}
%\label{figlabel}            % for cross-references
\end{center}
\end{figure}

\begin{figure}
\begin{center}
\centerline{\psfig{file=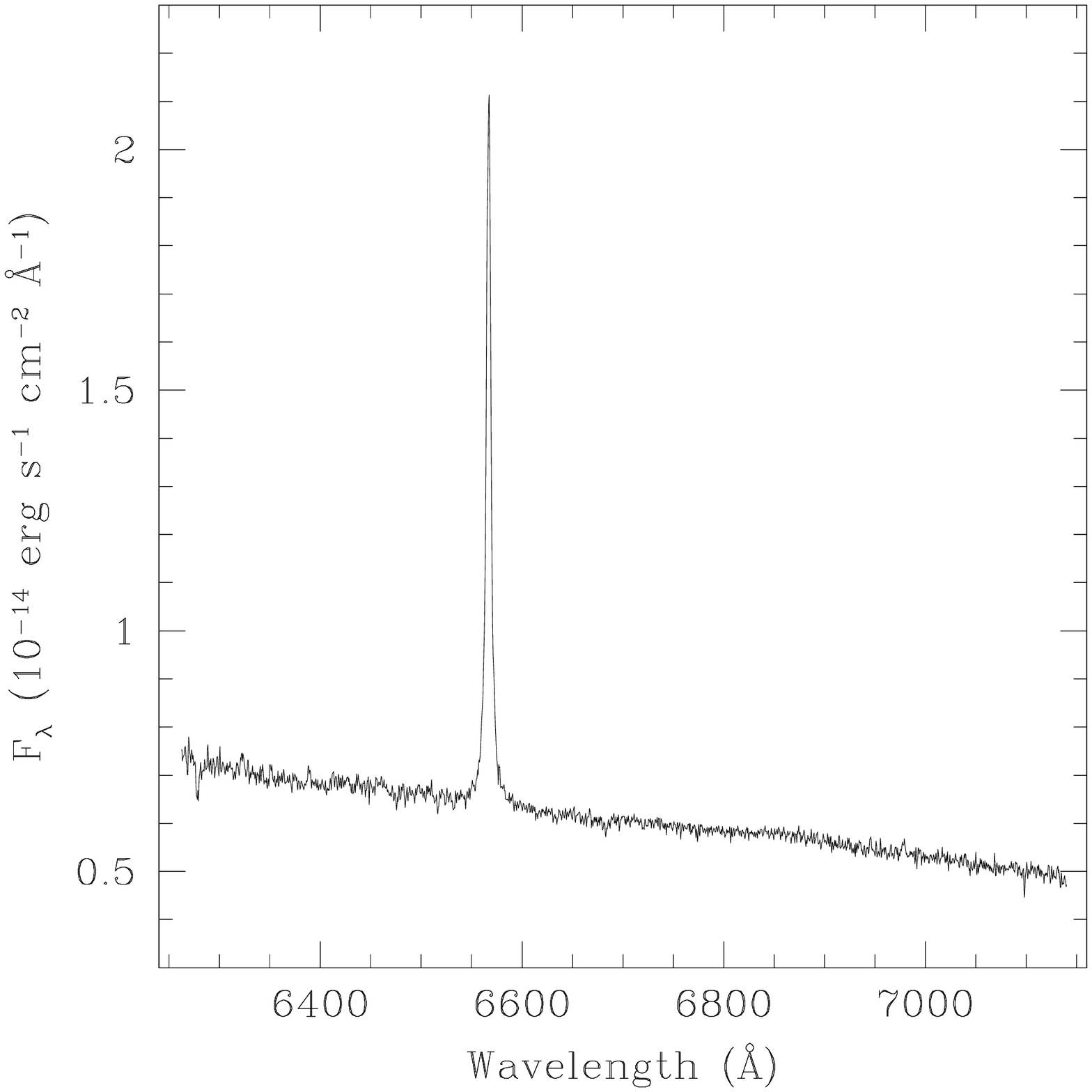,width=17cm}}
\caption{Flux-calibrated red spectrum of the optical counterpart 
of RX J0117.6$-$7330, taken on August 20, 1998 from the ANU 2.3m telescope 
at Siding Spring Observatory}
%\label{figlabel}            % for cross-references
\end{center}
\end{figure}

The most prominent feature in the blue spectral region is 
a strong H$\beta$ emission line: the equivalent width of the emission 
core, defined as in Dachs et al.\ (1981), is $= (-1.35 \pm 0.05)$ \AA; 
broader photospheric absorption wings are also present. 
H$\gamma$ is seen in absorption with a narrower and weaker emission 
core [EW $= (-0.27 \pm 0.02)$ \AA]. Narrow 
absorption is observed from He{\small{ I}} $\lambda 4388$ (EW $= 0.4$ \AA), 
He{\small{ I}} $\lambda \lambda 4471, 4472$ (EW $= 0.6$ \AA) 
and He{\small{ I}} $\lambda 4922$ (EW $= 0.3$ \AA). Figure 3 and Figure 4 
show the region of the blue spectrum (average of all four nights, 
normalised to the continuum) around H$\gamma$ and H$\beta$. Other 
weaker lines identified in the blue spectrum are listed in Table 1.

\begin{table}
\caption{Most prominent lines found in the optical spectrum of 
RX J0117.6$-$7330 (Balmer lines excluded)} 

\begin{tabular}{lr} \\ \hline
Line & EW (\AA)\\
\hline
He{\small{ II}} $\lambda 4200$  & $0.05 \pm 0.02$ \\
O{\small{ II}} $\lambda 4254$ & $0.10 \pm 0.02$ \\
O{\small{ II}} $\lambda 4317$ & $0.10 \pm 0.02$ \\
O{\small{ II}} $\lambda 4349$ & $0.15 \pm 0.05$ \\
O{\small{ II}} $\lambda 4367$ & $0.03 \pm 0.01$ \\
He{\small{ I}} $\lambda 4388$ & $0.40 \pm 0.10$ \\
O{\small{ II}} $\lambda \lambda 4415, 4417$ & $0.07 \pm 0.02$ \\
He{\small{ I}} $\lambda 4438$ & $0.08 \pm 0.02$ \\
He{\small{ I}} $\lambda \lambda 4471, 4472$ & $0.60 \pm 0.10$ \\
S{\small{ II}} $\lambda 4550$ & $-0.07 \pm 0.02$ \\
N{\small{ II}}/Si{\small{ III}} $\lambda 4553$ & $0.10 \pm 0.02$ \\
Si{\small{ III}} $\lambda 4568$ & $0.09 \pm 0.02$ \\
O{\small{ II}} $\lambda 4642$ & $0.15 \pm 0.02$ \\
O{\small{ II}} $\lambda \lambda 4649, 4651$ & $0.25 \pm 0.05$ \\ 
O{\small{ II}} $\lambda 4662$ & $0.08 \pm 0.02$ \\
O{\small{ II}} $\lambda 4676$ & $0.09 \pm 0.02$ \\
He{\small{ II}} $\lambda 4686$ & $0.10 \pm 0.02$ \\
He{\small{ I}} $\lambda 4713$ & $0.12 \pm 0.02$ \\
He{\small{ I}} $\lambda 4922$ & $0.30 \pm 0.05$ \\
S{\small{ II}} $\lambda 6386$ & $-0.20 \pm 0.02$ \\ 
He{\small{ I}} $\lambda 6678$ & $0.20 \pm 0.02$ \\
\hline
\end{tabular}
\end{table}

In the red spectral region (Figure 5)
the strongest emission line is H$\alpha$ (EW $= -16.0 \pm 1.0$ \AA); 
weaker emission is seen 
from S{\small{ II}} $\lambda 6386$ (EW $= -0.20 \pm 0.02$ \AA); 
He{\small{ I}} $\lambda 6678$ is seen in absorption 
(EW $= 0.20 \pm 0.02$ \AA).

Based on these features, the primary star can be identified as a B$0.5$\,IIIe, 
consistently with the results of Charles et al.\ (1996), 
and of Clarke et al.\ (1997).

\begin{figure}
\begin{center}
\centerline{\psfig{file=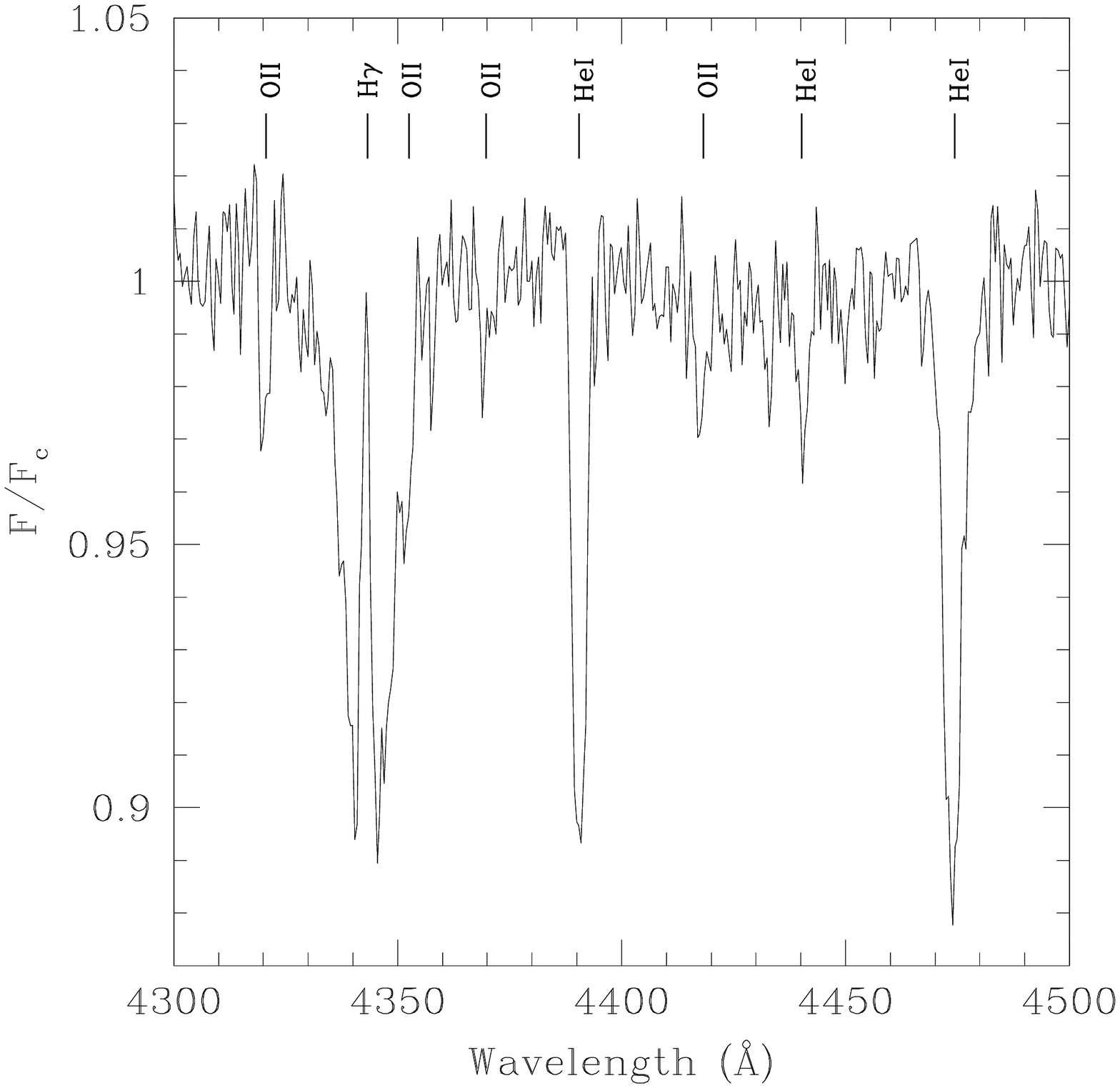,width=17cm}}
\caption{Portion of the coadded blue spectrum normalised to the continuum; 
some of the absorption lines listed in the text have been identified here.}
%\label{figlabel}            % for cross-references
\end{center}
\end{figure}

\begin{figure}
\begin{center}
\centerline{\psfig{file=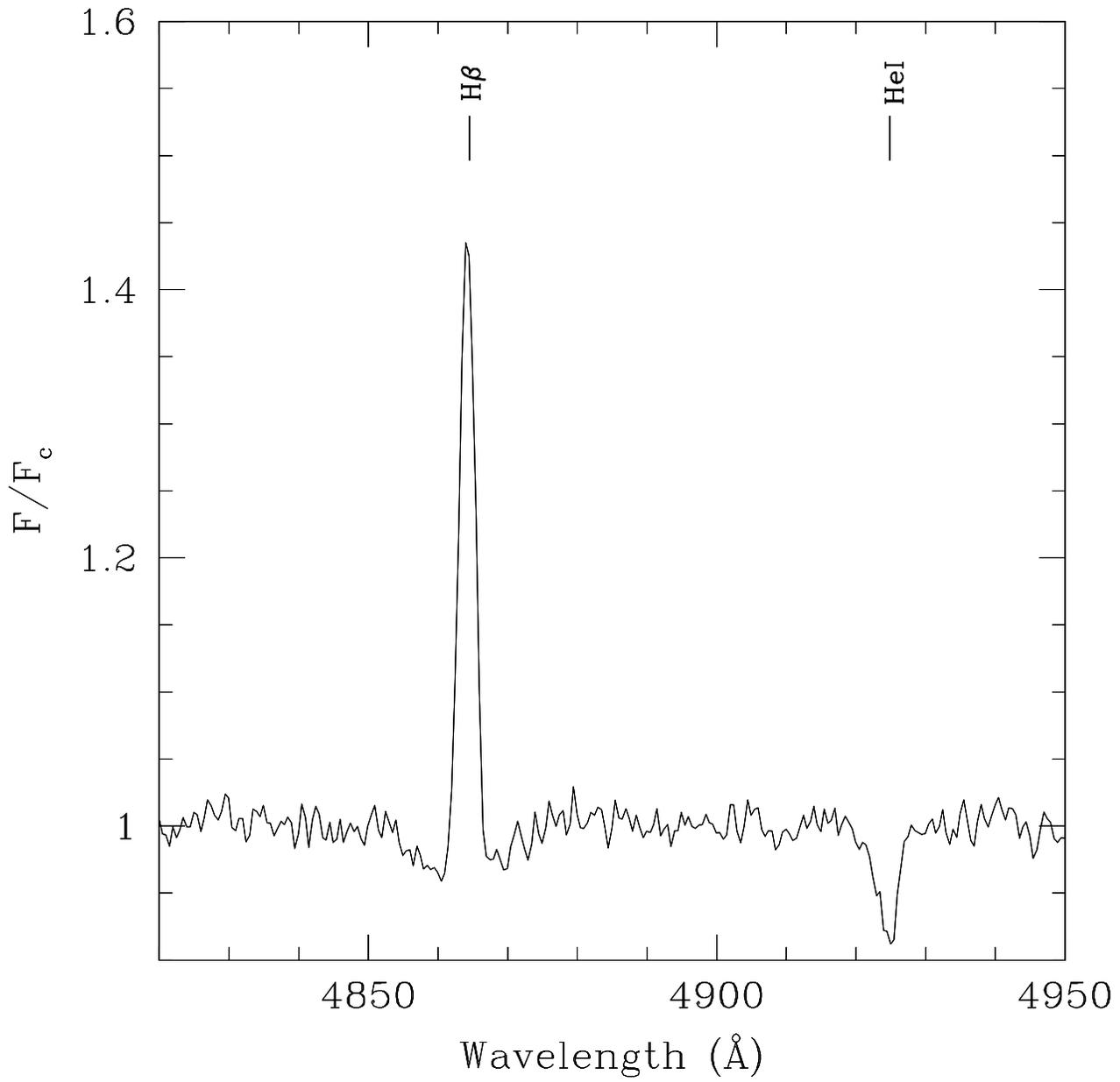,width=17cm}}
\caption{Portion of the coadded blue spectrum, normalised to the 
continuum. H$\beta$ and He{\small{ I}} $\lambda 4922$ are the main features.
}
%\label{figlabel}            % for cross-references
\end{center}
\end{figure}

\begin{figure}
\begin{center}
\centerline{\psfig{file=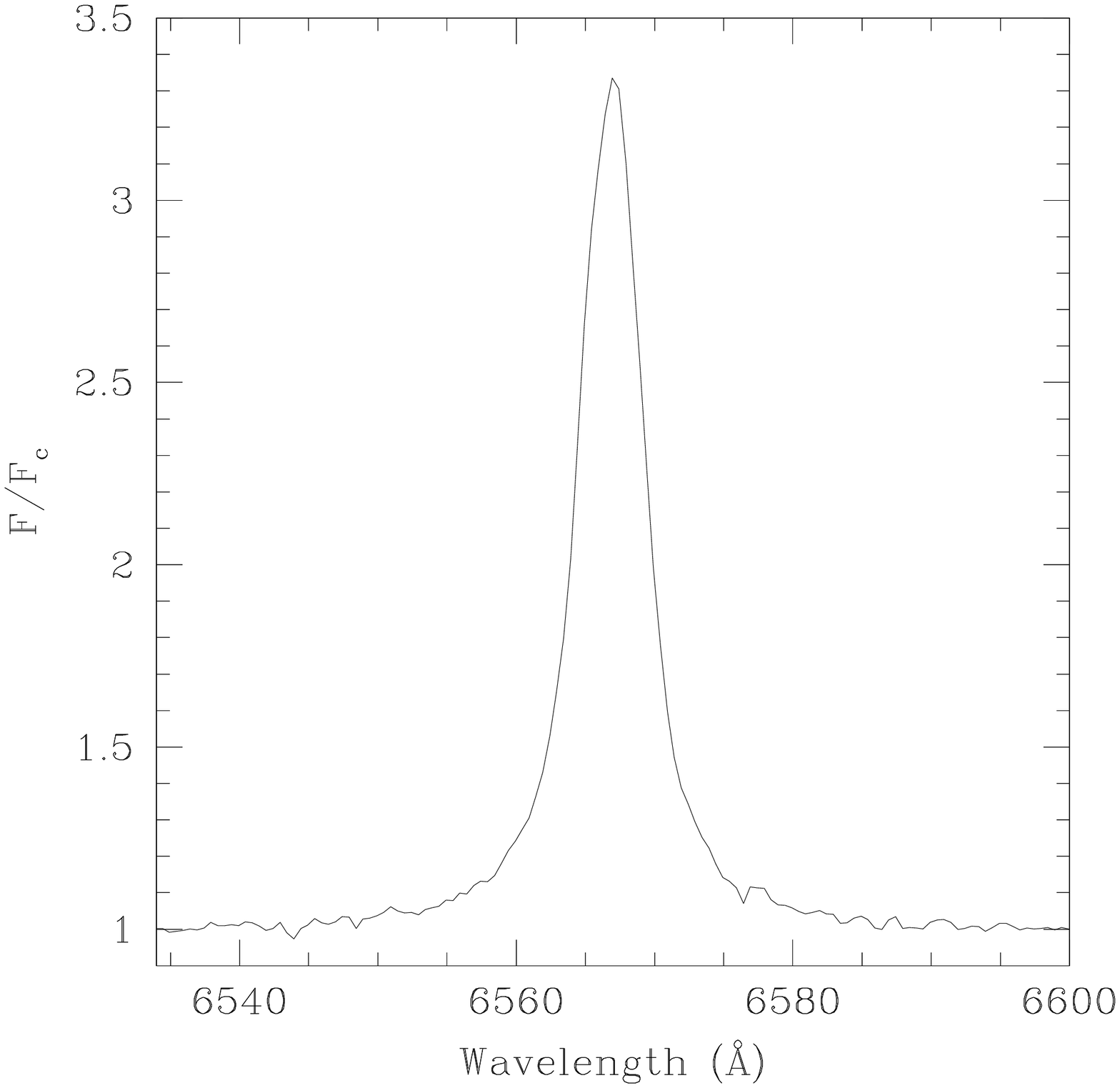,width=17cm}}
\caption{H$\alpha$ line profile, normalised to the continuum, in the 
coadded red spectrum. Balmer lines always appear single-peaked, 
and their profile 
is quite different from that expected for emission from a thin disk.}
%\label{figlabel}            % for cross-references
\end{center}
\end{figure}

\subsection{Photometry}

Photometric observations of the system were conducted 
from the SSO 40inch telescope; the detector used was a 
SITe $2048 \times 2048$ CCD. We obtain   
apparent magnitudes $B = 14.12 \pm 0.01$, $V = 14.19 \pm 0.01$, 
$R = 14.13 \pm 0.01$ and $I = 14.10 \pm 0.01$; no significant 
variations in the brightness of the star were observed during the run. 
Following Clarke et al.\ (1997) 
[see also van der Klis et al.\ (1992)], 
we have adopted a reddening of $E(B-V) = 0.08 \pm 0.01$; 
adopting also a distance modulus 
for the SMC of $d = 18.9$ (Feast 1991) we get absolute magnitudes 
$M_B = -5.02 \pm 0.04$, $M_V = -4.87 \pm 0.03$, $M_I = -4.86 \pm 0.02$.

We expect the Be star to appear redder than a non-Be star of similar 
temperature (Bessell 1993) because of the radiation from 
the circumstellar disk (colder than the star); Paschen continuum emission 
usually gives a particularly significant contribution in the near IR.
Assuming that the $B$ magnitude is the least 
affected by this additional contribution, and 
using the theoretical isochrones of Bertelli et al.\ (1994) in the range  
of metallicities $Z = 0.002 - 0.004$ (Bessell 1993) we can 
estimate a bolometric magnitude $M_{\mbox{{\footnotesize bol}}} = -7.4 \pm 0.2$
and an effective temperature 
$\log T_{\mbox{{\footnotesize eff}}} = 4.44 \pm 0.03$.
These values correspond to the giant phase of evolution for 
stars of mass $M_* = (18 \pm 2) M_{\odot}$, and are therefore consistent with 
the spectral identification of the optical counterpart of RX J0117.6$-$7330 
as a B$0.5$\,IIIe star. 

Using the results of Underhill et al.\ (1979), we can also infer a radius 
$R_* \simeq 10 R_{\odot}$, although values for the same 
spectral types determined by Popper (1980) are lower by $\sim 30$\%.

\subsection{Projected rotational velocity and radial velocity}

An interesting feature of our spectra is the small full width at half maximum 
(FWHM) of all the lines; the narrowest absorption lines are
He{\small{ I}} $\lambda 4388$ and He{\small{ I}} $\lambda 4922$, for 
both of which we calculate an average FWHM $= (210 \pm 20)$ \kms. 
The Doppler broadening of spectral lines is a function of the 
projected rotational velocity $v \sin i$. The FWHM of the 
He{\small{ I}} $\lambda 4471$ absorption line was used by 
Slettebak et al.\ (1975) as a parameter for a system of standard
rotational velocity stars. We determine an average 
FWHM $= (4.3 \pm 0.3)$ \AA\ for He{\small{ I}} $\lambda 4471$ 
in our spectra; correcting for the instrumental broadening 
(resolution $= 1.2$ \AA), we estimate a FWHM $= (4.1 \pm 0.3)$ \AA\ 
$= (275 \pm 20)$ \kms. Comparing this value with those listed in 
Slettebak et al.\ (1975) for the same spectral type, we estimate 
a projected rotational velocity $v \sin i = (145 \pm 10)$ \kms.

It is generally assumed (Hardorp \& Strittmatter 1970) 
that all Be stars are fast rotators with approximately the same 
rotational velocity, the observed velocity spread 
being due to orientation effects. The largest values of $v \sin i$ 
measured from 
line profiles are in the neighbourhood of $400$ \kms (Sletteback 1982). 
If we assume a true rotational velocity at the equator $v = (400 \pm 50)$ 
\kms, we infer an inclination angle $i = (21 \pm 3)\deg$.

An empirical correlation between the full width at half maximum 
of the emission component from H$\alpha$, its 
equivalent widths and the projected rotational velocity was derived 
by Dachs et al.\ (1986):
\\
\begin{equation}
\frac{{\mbox{FWHM(H$\alpha$)}}}{2} \left[\frac{{\mbox{EW(H$\alpha$)}}}
{-3 {\mbox{\AA}}}\right]^{1/4} -60 \simeq  
(v \sin i \pm 30) \quad {\mbox{\kms}}.
\end{equation}
\\
In this case, we measure a mean FWHM $= (6.1 \pm 0.2)$ \AA\ $= (280 \pm 10)$ 
\kms, and a mean EW $= (-16.0 \pm 1.0)$ \AA\ for H$\alpha$ (Figure 5).
This would lead to a projected rotational velocity $v\sin i = (150 \pm 30)$ 
\kms, in agreement with the more reliable value derived 
from the He{\small{ I}} $\lambda 4471$ absorption line.

It is reasonable to assume (Dachs et al.\ 1986) that the equivalent width 
of the H$\alpha$ emission line is proportional to the projected area of 
the disk orbiting the Be star in the equatorial plane; the disk is 
made of gas excreted from the star, and its outer radius is expected 
to increase during an active phase of the system. Using the empirical relation 
between H$\alpha$ equivalent width and disk radius given by 
Dachs et al.\ (1996), we derive $R_d = (3.4 \pm 0.1) R_*$, where $R_*$ is 
the radius of the Be star and $R_d$ is the radius at which optical depth 
equals unity for H$\alpha$ emission. As discussed in \S2.2, we can take 
$R_* = (10 \pm 3) R_{\odot}$ and $M_* = (18 \pm 2) M_{\odot}$. 

If the circumstellar disk were geometrically thin and in keplerian rotation, 
the emitting gas at its outer rim would have a projected rotational velocity 
\\
\begin{equation}
v_d \sin i = \left(\frac{GM_*}{R_d}\right)^{1/2} \sin i 
= (114 \pm 25) \quad {\mbox {\kms}}.
\end{equation}
\\
We would therefore expect to observe double-peaked line profiles for 
H$\alpha$ and H$\beta$ with peak-to-peak separations $\simeq 2v_d \sin i 
\simeq 230$ \kms (Smak 1981). This value corresponds to a separation 
$\Delta \lambda = 3.7$ \AA\ at H$\beta$, and 
$\Delta \lambda = 5.0$ \AA\ at H$\alpha$, well discernible with our $1.2$ 
\AA\  
resolution. We observe that both H$\alpha$ and H$\beta$ emission line profiles 
are always symmetrical and single-peaked: this suggests 
that the circumstellar gas is not 
confined to a thin disk in the equatorial plane, 
but may form a thick torus or an envelope which extends 
to the polar regions of the stellar atmosphere. Alternatively, absence 
of double peaks could be due to non-keplerian motion in the outer disk, 
where radial outflows can dominate over rotation, or 
to a much larger disk radius [cf.\ the model proposed by 
Poeckert \& Marlborough 
(1979)]. As shown in Dachs et al.\ (1996), the 
H$\alpha$ emission lines from Be stars are almost always single-peaked 
for values of EW, FWHM and $v \sin i$ 
similar to those measured in this system.

\begin{figure}
\begin{center}
\centerline{\psfig{file=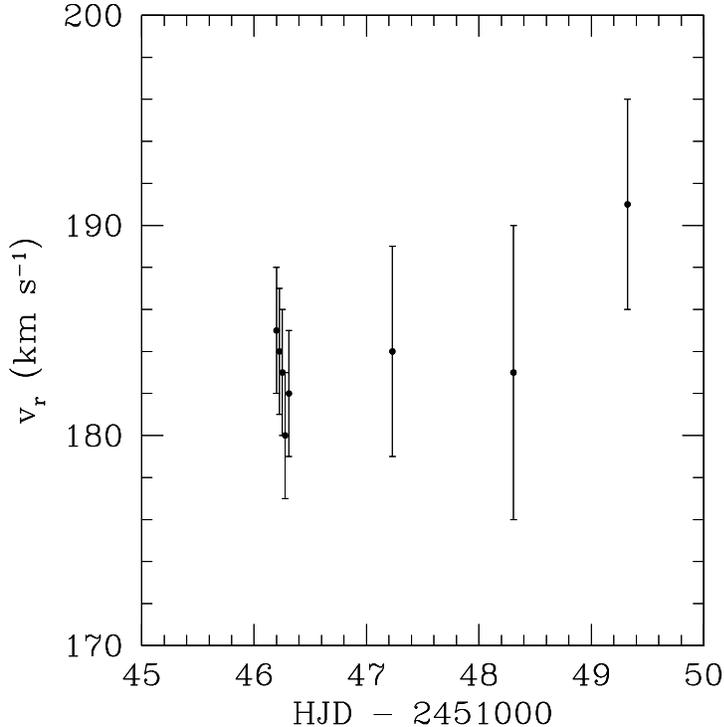,width=10cm}}
\caption{Radial velocity of the optical counterpart of RX J0117.6$-$7330.}
%\label{figlabel}            % for cross-references
\end{center}
\end{figure}

The projected radial velocity of the system was determined by measuring 
the central position (using a Gaussian fit) of H$\alpha$, H$\beta$, 
He{\small{ I}} $\lambda 4388$ and He{\small{ I}} $\lambda 4922$ in 
each spectrum (two or three consecutive 600s spectra were averaged 
together to increase the S/N); the values found are 
plotted in Figure 6.
Although the variations in the measured radial velocity may be 
due to the orbital motion, the data are insufficient to determine 
the orbital period or the 
radial velocity amplitude from these data, or the eccentricity 
of the orbit (it can be $e \simgt 0.3$ in Be/X-ray binary 
systems). All known Be/X-ray binaries have orbital periods $P \simgt 15$ days, 
with periods of hundreds of days in some cases (van den Heuvel \& Rappaport 
1987). The average systemic velocity over the time of our spectral 
observations is $\gamma = (184 \pm 4)$ \kms, confirming that 
the system is located in the SMC as suggested by Clarke et al.\ (1997).

\section{Conclusions}

We have presented spectroscopic and photometric observations of the 
optical counterpart of the X-ray transient RX J0117.6$-$7330 in 
the Small Magellanic Cloud, currently in an inactive state. 
The data allow us to identify the primary star as a B$0.5$\,IIIe star, 
of mass $M_* = (18 \pm 2) M_{\odot}$, radius $R_* = (10 \pm 3) R_{\odot}$, 
effective temperature 
$T_{\mbox{{\footnotesize eff}}} = (2.75 \pm 0.20)\times 10^4$ K and 
bolometric magnitude $M_{\mbox{{\footnotesize bol}}} = (-7.4 \pm 0.2)$ mag.
We have derived an inclination angle of the binary system 
$i = (21 \pm 3)\deg$ and an average projected radial velocity
$\gamma = (184 \pm 4)$ \kms; we have also measured the 
equivalent widths of the main optical lines. Further observations 
at different orbital phases are necessary to determine the binary period and 
the mass function of the compact object. Comparison between this set 
of data and observations taken during an X-ray active state (phase of 
enhanced mass loss from the Be star) may be useful to determine 
the orbital parameters of the binary systems and to improve the current 
models of mass transfer in HMXBs.

%
% Add as many section titles/contents as required.
%
% If you have subsections then use the
% \subsection{SUBSECTION TITLE}
% command and if you have subsubsections then use the
% \subsubsection{SUBSUBSECTION TITLE}
% command.  To use these commands, 
% first remove the % from the start of the line.

% It is preferable to embed your figures in the text. 
% One way to do this is to use the psfig style file and use the following
% commands to include the figures:

% \begin{figure}
% \begin{center}
% \psfig{file=filename.ps,height=10cm}
% \caption{Write your figure caption here.}
% \label{figlabel}            % for cross-references
% \end{center}
% \end{figure}

% To use the above commands, first remove the % from the beginning of
% the lines and then fill in your own values etc as appropriate.

% Tables
% Please consult previous issues of PASA
%  to see how tables are to be formatted.

\section*{Acknowledgements}

I would like to thank Kinwah Wu who did some of the 
photometric observations at the SSO 40inch telescope; 
thanks also to Stefan Keller and Mike Bessell for their 
useful comments and suggestions about Be stars.

% Place acknowledgements here. Omit above \section command if there
% are no acknowledgements.

\section*{References}

% PASA uses the same conventions as ApJ for journal abbreviations.  Sample
% references are as follows. 
% Please follow the same format for your references.

%\reference Author, A.B. 1990 PASA 7, 2, 350

% for a journal article, or

% \reference Author, A.B. 1990 in This Is A Book Title, ed. Editor, C.D.,
% This Is A Publishers Name, 437

% for a book.

\reference Bertelli, G., Bressan, A., Chiosi, C., Fagotto, F., \& Nasi, E. 
1994, A\&AS, 106, 275
\reference Bessell, M. S. 1993, in Precision Photometry, ed. D. Kilkenny,
E. Lastovica \& J. W. Menzies (Cape Town: The Observatory), 227
\reference Charles, P. A., Southwell, K. A.,  \& O'Donoghue, D. 1996, 
IAU Circ.\ 6305
\reference Clarke, G. W., Remillard, R. A., \& Woo, J. W. 1996, IAU Circ.\ 6282
\reference Clarke, G. W., Remillard, R. A., \& Woo, J. W. 1997, ApJ, 474, L111
\reference Dachs, J. et al.\ 1981, A\&AS, 43, 427
\reference Dachs, J., Hanuschik, R., Kaiser, D., \& Rohe, D. 1986, 
A\&A 159, 276
\reference Feast, M. W. 1991, in The Magellanic Clouds, IAU Symposium 148, 
ed. R. Haynes \& D. Milne (Dordrecht: Kluwer), 1
\reference Hardorp, J., \& Strittmatter, P. A. 1970, in Stellar Rotation, 
ed. A. Slettebak (Dordrecht: Reidel Publishing Company), 48
\reference Kahabka, P., \& Pietsch, W. 1996, A\&A, 312, 919
\reference Poeckert, R., \& Marlborough, J. M. 1979, ApJ, 233, 259
\reference Popper, D. M. 1980, ARA\&A, 18, 115
\reference Slettebak, A. 1982, ApJS, 50, 55
\reference Slettebak, A., Collins, G. W., Boyce, P. B., White, N. M., 
\& Parkinson, T. D. 1975, ApJS, 29, 137
\reference Smak, J. 1981, Acta Astron., 31, 395
\reference Underhill, A. B., Divan, L., \& Pr\'{e}vot-Burnichon, M.-L. 1979, 
MNRAS, 189, 601
\reference van den Heuvel, E. P. J., \& Rappaport, S. 1987, in 
Physics of Be Stars, ed. A. Slettebak \& T. P. Snow (Cambridge: University 
Press), 291
\reference van Paradijs, J., \& McClintock, J. E. 1995, 
in X-Ray Binaries, ed. W. H. G. Lewin, J. van Paradijs, \& 
E. P. J. van den Heuvel (Cambridge Astrophysics Series, 
Cambridge University Press), 536
\reference van der Klis et al.\ 1992, A\&A, 106, 339

% Add as many references as required.

\end{document}